# Scheduling in a random environment: stability and asymptotic optimality*

U. Ayesta[1,2], M. Erausquin[1,3], M. Jonckheere[4], I.M. Verloop[1]

[1]BCAM – Basque Center for Applied Mathematics, Derio, Spain
[2]IKERBASQUE, Basque Foundation for Science, Bilbao, Spain
[3]UPV/EHU, University of the Basque Country, Bilbao, Spain
[4]CONICET, Buenos Aires, Argentina

*Abstract*—We investigate the scheduling of a common resource between several concurrent users when the feasible transmission rate of each user varies randomly over time. Time is slotted and users arrive and depart upon service completion. This may model for example the flow-level behavior of end-users in a narrowband HDR wireless channel (CDMA 1xEV-DO). As performance criteria we consider the stability of the system and the mean delay experienced by the users. Given the complexity of the problem we investigate the fluid-scaled system, which allows to obtain important results and insights for the original system: (1) We characterize for a large class of scheduling policies the stability conditions and identify a set of maximum stable policies, giving in each time slot preference to users being in their best possible channel condition. We find in particular that many opportunistic scheduling policies like Score-Based [8], Proportionally Best [1] or Potential Improvement [4] are stable under the maximum stability conditions, whereas the opportunistic scheduler Relative-Best [9] or the $c\mu$-rule are not. (2) We show that choosing the right tie-breaking rule is crucial for the performance (e.g. average delay) as perceived by a user. We prove that a policy is asymptotically optimal if it is maximum stable *and* the tie-breaking rule gives priority to the user with the highest departure probability. We will refer to such tie-breaking rule as *myopic*. (3) We derive the growth rates of the number of users in the system in overload settings under various policies, which give additional insights on the performance. (4) We conclude that simple priority-index policies with the myopic tie-breaking rule, are stable and asymptotically optimal. All our findings are validated with extensive numerical experiments.

## I. INTRODUCTION

Next generation wireless networks are expected to support a wide variety of data services. Due to fading and interference effects, for each user, the quality of a downlink channel, and hence its transmission rate, fluctuates over time. This has triggered a large amount of work aiming at understanding the performance of channel-aware scheduling policies. It is by now accepted that so-called "opportunistic schedulers" have many desirable properties (see for example [9]). A policy is called opportunistic if it takes advantage of the channel fluctuations by serving a user whose channel condition is in a good state with respect to its own statistical behavior. With the objective of minimizing mean users' delay, there arises a key tradeoff in the design of scheduling mechanisms between making full use of the opportunistic gains, and prioritizing users having small residual service sizes.

Broadly speaking, researchers have explored scheduling in wireless systems both at the packet level and at the flow level. In packet-level models it is typically assumed that there exists a finite number of permanent users. The focus of the scheduler is on the number of packets in the queue of each user. We refer for example to [34], [2], [33], [17], [3], [25], [29] for this line of research. In a flow-level model instead, users arrive randomly to the system and leave after receiving their finite-sized service demands. This allows to capture the performance as perceived by the end-users, see for example [8], [21], [9], [26], [22], [1], [4], [30]. For surveys on flow-level modeling we refer to [24] and [10]. In [23], hybrid models are studied.

The performance evaluation and optimization of wireless networks at the flow level has proved to be extremely challenging. One of the most successful approaches has been the so-called time-scale separation argument (see [9], [11], [12], [29], [1], [30]) where it is assumed that at the flow scale the dynamics of the channel fluctuations can be averaged out. Under this time-scale assumption, it was shown in [11] that any utility-based scheduling policy is stable in a flow-level model. The authors of [1] make the same assumption when they discuss rate-stability for priority-index policies. In the context of optimal control, in [29], [30] scheduling mechanisms were introduced and evaluated. In [4] optimal control is studied without the time-scale separation assumption. The Lagrangian-relaxation method allowed the authors of [4] to construct the Potential Improvement (PI) scheduling policy, which is optimal for a relaxed optimization problem. In addition, several other policies have been proposed and numerically investigated in the literature, among others the Proportional Fair [14] discipline, the Score-Based (SB) algorithm [8], the Relative Best (RB) scheduler [6] and Proportionally Best (PB) [1].

To sum up, without a time-scaling separation argument, which is a rather strong assumption, the performance of opportunistic schedulers, regarding stability and performance perceived by the users, is not well understood. In order to gain better insight into the latter issue, in this paper we will study a flow-level model without the time-scale separation assumption.

*Research partially supported by grant MTM2010-17405 of the MICINN (Spain) and grant PI2010-2 of the Basque Government (Department of Education and Research). Martin Erausquin's PhD. is supported by grants ECO2008-00777 and FPU AP2008-02014, both of the MEC (Spain).

More precisely, we assume that data users arrive randomly in time and have a finite amount of data to download. Time is slotted and the quality of the channel condition of each user varies per time slot. In every time slot at most one user may be served. We are interested in stability and optimization of the system. Given the complexity of the problem we first prove convergence of the fluid-scaled system towards a unique fluid limit. We note that the precise characterization of the fluid limit involves averaging phenomena of the scaled system which is not grasped by the usual description of weak fluid limits.

The fluid-limit description allows us to obtain several important results and insights for the original wireless system. First of all, we characterize the maximum stability conditions (the weakest possible conditions on the traffic parameters such that there exists a scheduling policy that makes the system positive recurrent) and show that the set of policies that are stable under the maximum stability condition have a very simple characterization: whenever there are users present that are currently in their best channel condition, only such users are served. These policies will be referred to as Best-Rate (BR) policies. Such a characterization was previously given for rate stability [1], but, to the best of our knowledge, stochastic stability was still an open issue.

Second, for a large class of scheduling policies we determine the exact stability conditions and conclude that many known opportunistic scheduling policies like SB, PB, or PI are stable under the maximum stability conditions, whereas the opportunistic scheduler RB or $c\mu$-rule are not.

Third, we demonstrate the importance for the choice of the tie-breaking rule when the goal is to optimize the performance. Until now, the literature proposed to break ties at random, see for example [6], [8], [9], [1]. We instead propose to give priority to the user with highest instantaneous departure probability when there are multiple users in their best channel conditions, which we refer to as the myopic tie-breaking rule. We prove that BR policies with the myopic tie-breaking rule are asymptotically fluid optimal and our numerical experiments further illustrate that the myopic tie-breaking rule significantly improves the performance. This in turn shows that simple priority-index policies that balance opportunistic gains with size-based information, will be both maximum stable and asymptotically optimal.

Fourth, our convergence result allows to compare the performance of the various policies in an overload setting. More precisely, we determine the growth rates of the number of users in the various classes and find that BR policies with a myopic tie-breaking rule minimize the total growth rate.

The paper is organized as follows. In Section II we present the model. In Section III we introduce the scheduling policies of interest and define their tie-breaking rule. In Section IV we derive fluid limits for a large class of policies. This allows to obtain our stability results as presented in Section V. In Section VI we characterize asymptotically optimal policies, both in normal regime and in overload, and discuss the importance of the tie-breaking rule. In Section VII we perform numerical experiments to validate our theoretical findings.

## II. Model Description

We consider a time-slotted system serving one user in each time slot. This models for instance a CDMA 1xEV-DO system as explained in Remark 1. There are $K$ classes of users, and in each time slot the number of class-$k$ users arriving to the system, $A_k$, follows an i.i.d. sequence of random variables, with $\mathbb{E}(A_k) = \lambda_k$ and $\mathbb{E}(A_k^2) < \infty$. For each user the departure probability varies over time as the quality of the channel is changing from slot to slot. The quality of the channel (or state of the channel) for a class-$k$ user is modeled as an i.i.d. sequence of random variables taking values in the finite set $\mathcal{N}_k := \{1, 2, \ldots, N_k\}$. For each time slot we let $q_{k,n}$ denote the probability that a class-$k$ user is in channel state $n \in \mathcal{N}_k$. Associated with channel state $n$ is a departure probability $\mu_{k,n}$. This can be used for instance to model a system in which the service requirements are geometric (see Remark 1). Without loss of generality we assume that the channel conditions are ordered such that $0 \leq \mu_{k,1} \leq \mu_{k,2} \leq \cdots \leq \mu_{k,N_k} \leq 1$, and $q_{k,N_k}\mu_{k,N_k} \neq 0, \ \forall \ k$. The channel condition of a class-$k$ user is independent of the channel conditions of all the other users and of the channel quality history.

In each time slot $t$, a scheduler/policy $f$ decides which user is served. Because of the Markov property of the system, we focus on policies that base decisions on the current number of users present in the various classes and on their current channel states. For a given scheduling policy $f$, let $X_k^f(t)$ denote the number of class-$k$ users in the various classes at time slot $t$ and $X^f(t) = (X_1^f(t), \ldots, X_K^f(t))$. Since the channel conditions are i.i.d. and independent of the process $X^f(\cdot)$, the process $X^f$ is Markov and, in addition, for the modeling it is sufficient to focus on the Markovian description in terms of the number of users in each *class*, $X^f(\cdot)$, instead of the number of users in each *channel state*.

Let us introduce some more notation. We denote by $|x|$ the $l_1$ norm of a vector $x$. The notation $x \leq y$ is used for the coordinate-wise ordering: $x_i \leq y_i, \forall i$. Finally, we denote by u.o.c. the uniform convergence on compact sets.

*Performance criteria:* Our performance criteria are stability and long-run average number of users. We use the following definition for stability:

**Definition 1.** *A scheduling policy $f$ is stable if the process $X^f$ is positive recurrent.*

Because of the time-varying channel conditions the system is not work-conserving and hence it depends strongly on the employed scheduling policy whether the system can be made stable. We define the *maximum stability conditions* as the conditions on the traffic inputs such that there exists a policy that can make the system stable. A *maximum stable policy* is a policy that is stable under the maximum stability conditions. From the performance point of view it is therefore of crucial importance to design a scheduler that is maximum stable.

Besides stability, another important performance measure is the long-run time-average holding cost,

$$\limsup_{T \to \infty} \frac{1}{T} \sum_{k=1}^{K} \sum_{t=0}^{T} c_k \mathbb{E}(X_k^f(t)), \qquad (1)$$

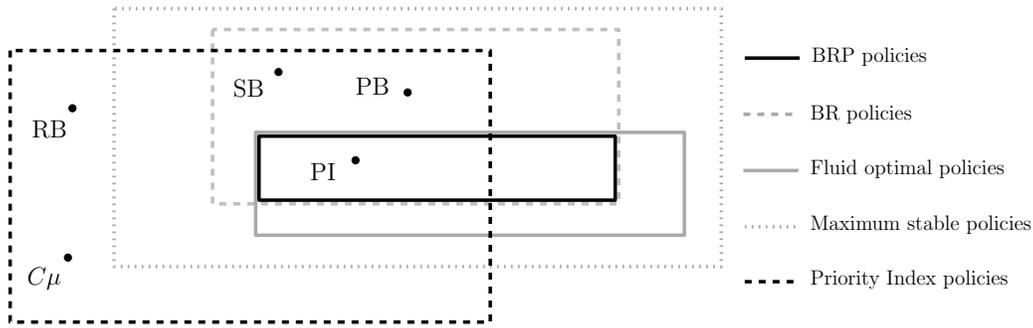

Fig. 1. Classification of schedulers.

with $c_k > 0$ the holding cost incurred per time slot for having a class-$k$ user in the system. When $c_k = 1, \forall k$, this is equivalent to minimizing the mean sojourn time (cf. Little's law).

**Remark 1 (Modeling of a wireless data network).** *Our model, even though simple, captures some of the key properties of wireless communication systems. Time is slotted, as is the case in the CDMA 1xEV-DO [5] and the OFDM-based LTE systems [31]. The available transmission rate of each user fluctuates due to fading effects, and as a consequence, it varies from one slot to another. We note that in real systems the number of feasible transmission rates is finite (see [5]).*

One may classify the users into different classes based on their applications or traffic conditions for example. Let the service requirement (in bits) of a class-$k$ user be a geometric random variable denoted by $B_k$, and let $\mathbb{E}(B_k)$ denote its expectation. Let $\Delta$ denote the amount of bits transferred in one slot under the current channel condition. Note that in practice $\Delta$ will vary from slot to slot depending on the channel condition, and the allocation. The probability that a user leaves the system is approximately $\mathbb{P}(b \leq B_k \leq b + \Delta | B_k > b) \approx \Delta / \mathbb{E}(B_k)$, which does not depend on the attained service $b$ (memoryless property of the geometric distribution). This expression becomes asymptotically exact as the ratio $\Delta / \mathbb{E}(B_k)$ goes to 0. Hence, this is the case if the mean service requirement (in bits) of a user is very large compared to the amount of bits that can be served in one slot. Let $s_{k,n}$ denote the transmission rate (in bits per second) of a class-$k$ user when the channel state is $n$. For the CDMA 1xEV-DO system, the amount of bits transferred in one slot is $\Delta = s_{k,n} \cdot t_c$. Hence, the departure probability of a class-$k$ user under channel condition $n$ can be approximated by

$$\mu_{k,n} := \frac{s_{k,n} \cdot t_c}{\mathbb{E}(B_k)}, \qquad (2)$$

where $t_c$ is the length of the slot (for example $t_c = 1.67ms$ in the CDMA 1xEV-DO system). In Section VII we perform numerical experiments with departure probabilities as obtained from a practical setting using (2).

**Remark 2 (Modeling an OFDM system).** *A natural extension to our modeling framework will be to allow that in every slot multiple users can be served in parallel as it happens to be in the OFDM-based (3GPP LTE) system. In such a system there are $M$ subcarriers and a subcarrier can be assigned at most to one user at a time. As reported in the literature, in an OFDM system a user can experience a deep fading in one subcarrier, while on the same subcarrier another user could be in good condition [32]. Under i.i.d. assumptions it is expected that for the metrics under consideration in this paper it is of no interest to serve multiple classes in parallel, that is, in every slot all the sub-carriers will be assigned to only one of the classes. However, serving multiple classes in parallel could be though of interest for other metrics like fairness, an issue that even though not studied here, definitely deserves a thorough investigation.*

## III. POLICIES

In this section we introduce scheduling policies that will be used throughout the paper. Most of these policies are opportunistic (with the exception of the $c\mu$-rule), meaning that they take advantage of channel fluctuations by serving a user whose channel condition is currently in a good state, in some sense, with respect to its own statistical behavior.

We first introduce priority-index policies, which are very popular due to their simplicity from an implementation point of view. A priority-index policy is characterized by an index function that assigns an index to each user based solely on its class and its current state.

**Definition 2 (Priority-index policy).** *In every time slot, a user that has the highest index among all users present is served.*

Priority-index policies might need to be augmented with a suitable *tie-breaking rule*. Such a rule refers to the strategy adopted when there is a tie on the highest index value. A tie means that there are several users present having the highest index value, but these users belong to *different* classes. In the literature, most of the papers specify to break ties at random (see for example [8], [6], [1]). We define the *myopic tie-breaking rule* as the rule that among the users with the highest index, it selects the one with highest value for $c_k \mu_{k,N_k}, \forall k$ (the $c_k$'s refer to the holding cost introduced in Section II.) One of our main contributions will be that the choice for the tie-breaking rule is crucial for the performance of the system and that the myopic tie-breaking rule is close to optimal (this will be further developed in Sections VI and Section VII).

In [8] the Score-Based (SB) policy is introduced. SB is a priority-index policy where the index value of a class-$k$ user in state $n$ is given by $\sum_{\tilde{n}=1}^{n} q_{k,\tilde{n}}$, and ties are broken at random. In [4] the Potential Improvement (PI) policy is introduced. PI is a priority-index policy with index value $c_k\mu_{k,n}/\sum\limits_{\tilde{n}>n} q_{k,\tilde{n}}(\mu_{k,\tilde{n}} - \mu_{k,n})$, and the tie-breaking rule is the myopic tie-breaking rule. An important subset of the priority-index policies are the so-called weight-based policies.

**Definition 3 (Weight-based policy).** *A priority-index policy with index function $\omega_k \mu_{k,n}$. Here $\omega_k$ denotes a class dependent weight.*

Important examples of weight-based policies are: the $c\mu$-rule ($\omega_k = c_k$, with $c_k$ the holding cost), Relative Best (RB) [6] ($\omega_k = 1/\sum_{n=1}^{N_k} q_{k,n}\mu_{k,n}$), and Proportionally Best (PB) [1] ($\omega_k = 1/\mu_{k,N_k}$). For all these policies, ties are broken at random.

It will be convenient to define the following two classes of policies, which play an important role in the results on the stability analysis and asymptotic optimality.

**Definition 4 (Best Rate (BR) policies).** *The BR policies are such that whenever there are users present that are currently in their best channel condition, i.e., in state $N_k$, such a user is served.*

**Definition 5 (Best Rate Priority (BRP) policies).** *The BRP policies are BR policies with a myopic tie-breaking rule.*

As a consequence of our main results, we will obtain that the classes of policies BR and BRP have desirable properties: In Section V we prove that any BR policy is stable under the maximum stability conditions and in Section VI we derive that BRP policies are asymptotically optimal.

In Figure 1 we have summarized the various (classes of) policies. Note that SB, PB and PI are BR policies. This follows since the highest possible index value is 1 for SB and PB, and $\infty$ for PI, and these indices can only be obtained whenever a user is in its best possible channel condition. RB and the traditional $c\mu$ rule (i.e., giving in each time slot strict preemptive priority to the user having the highest $c_k\mu_{k,n}$) however do not belong to BR policies since, depending on the set of parameters, the index value of a class-$k$ user in state $n \neq N_k$ might be larger than the index value of a class-$l$ user in state $N_l$. Remark also that PI is the only BRP policy.

## IV. FLUID LIMITS AND CONVERGENCE

In this section we study fluid-scaling limits for a large class of policies. Fluid scaling or time-space scalings, corresponding to "zooming out" the trajectories, have been used extensively to study stochastic processes with complex dynamics [16]. The limiting processes are usually much simpler to describe while they provide crucial insights on the behavior of the non-scaled version of the process. In particular, the convergence results will allow us to prove crucial results on stability and optimality of schedulers for the stochastic system.

The fluid scaling consists in studying a sequence of systems indexed by $r$, i.e., for a given policy $f$ we let $X_k^{f,r}(t)$ denote the number of class-$k$ users at time $t$ when the initial state equals $X_k^r(0) = rx_k(0)$, $k = 1, \ldots, K$, with $r \in \mathbb{N}$, and $X^{f,r}(t) = (X_1^{f,r}(t), \cdots, X_K^{f,r}(t))$. We are then interested in the fluid-scaled processes $Y_k^{f,r}(t) := \frac{X_k^{f,r}(\lfloor rt \rfloor)}{r}$, $t \geq 0$, $k = 1, \ldots, K$, with $Y^r(0) = x(0)$. We can write

$$Y_k^{f,r}(t) = x_k(0) + \frac{1}{r}\sum_{s=1}^{\lfloor rt \rfloor} A_k(s) - \frac{1}{r}\sum_{n=1}^{N_k} S_{k,n}(T_{k,n}^{f,r}(rt)), \quad (3)$$

where $T_{k,n}^{f,r}(t)$ is defined as the cumulative amount of time that was spent on serving class $k$ in state $n$ during the interval $(0, t]$ and $S_{k,n}(t)$ denotes the total number of class-$k$ users that have been completed while receiving service for a total duration of time $t$ when being in state $n$.

In order to derive stability and fluid optimality results, we will be interested in limits of the fluid-scaled process. In Section IV-A we will characterize a generic description of weak fluid limits (usually not unique) of Equation (3), following the same reasoning as in [15]. In Section IV-B we focus instead on a special class of policies for which we can prove convergence in probability towards a *unique* limit, which will be referred to as the strong fluid limit. (In [20] similar is done but only for a subset of the state space.) We discuss the differences between weak and strong fluid limits in more detail in Remark 5, after having introduced formally both concepts.

### A. Convergence towards weak fluid limits

From (3), we obtain the following result that describes the generic characterization of weak fluid limits for a given policy $f$. The lemma will allow to determine maximum stable policies (Theorem V.2) and to characterize asymptotically optimal policies (Section IV).

**Lemma IV.1.** *For almost all sample paths $\omega$ and any sequence $r_k$, there exists a subsequence $r_{k_l}$ such that for all $k = 1, 2, \ldots, K$, $n = 1, 2, \ldots, N_k$, and $t \geq 0$,*

$$\lim_{l \to \infty} Y_k^{f,r_{k_l}}(t) = y_k^f(t), \quad u.o.c., \quad and \quad (4)$$

$$\lim_{l \to \infty} \frac{T_{k,n}^{f,r_{k_l}}(t)}{r_{k_l}} = \tau_{k,n}^f(t), \quad u.o.c.,$$

*with $(y_k^f(\cdot), \tau_{k,n}^f(\cdot))$ a continuous function. In addition,*

$$y_k^f(t) = x_k(0) + \lambda_k t - \sum_{n=1}^{N_k} \mu_{k,n}\tau_{k,n}^f(t), \quad (5)$$

*$y_k^f(t) \geq 0$, $\tau_{k,n}^f(0) = 0$, $\sum_{k,n} \tau_{k,n}^f(t) \leq t$, and $\tau_{k,n}^f(\cdot)$ are non-decreasing and Lipschitz continuous functions.*

**Proof:** The proof follows similarly to that of [13, Proposition 4.12]. Define $A_k(t)$ as the number of class-$k$ users that arrive in time slot $t$. We note that $A_k(1), A_k(2), \ldots$, are independent and distributed according to $A_k$, with $\mathbb{E}(A_k) = \lambda_k$. Since $\mu_{k,n}$ is the probability of completing a class-$k$ user when it receives service while being at state $n$, by the law of large numbers, we obtain that, almost surely,

$$\lim_{r \to \infty} \frac{1}{r}\sum_{s=1}^{\lfloor rt \rfloor} A_k(s) = \lambda_k t, \quad and \quad \lim_{r \to \infty} \frac{1}{r} S_{k,n}(rt) = \mu_{k,n} t. \quad (6)$$

Since $T_{k,n}^{f,r}(t)$ denotes the cumulative amount of time spent on serving class-$k$ users in state $n$ in time interval $[0,t]$, we get

$$\frac{T_{k,n}^{f,r}(rt)}{r} - \frac{T_{k,n}^{f,r}(rs)}{r} \leq t - s, \quad \text{for every } t \geq s,$$

i.e., $\overline{T}_{k,n}^{f,r}(t) := T_{k,n}^{f,r}(rt)/r$ is Lipschitz continuous. Therefore, by the Arzela-Ascoli theorem [27] we obtain that, for almost every sample path $w$ and for any subsequence $r_k$, there exists a subsequence $r_{k_l}$ of $r_k$ such that $\lim_{l\to\infty} T_{k,n}^{f,r_{k_l}}(rt)/r = \tau_{k,n}^f(t)$, u.o.c.. Now, using Equations (3) and (6), it follows that $\lim_{l\to\infty} Y_k^{f,r_{k_l}}(t) = y_k^f(t)$, with $y^f(t)$ as given in (5). $\quad\square$

We can now give our definition of weak fluid limits.

**Definition 6.** *We call the processes $\tau^f(t)$ and $y^f(t)$ (as obtained in Lemma IV.1) weak fluid limits for policy $f$.*

Note that, in general, these fluid limits can be different depending on the sample path and the subsequence considered. A policy is said to have a *unique fluid limit* if, for all sample paths and all subsequences, the weak fluid limits coincide.

### B. Convergence towards a strong fluid limit

In this subsection we will determine unique fluid limits for a special class of policies. More precisely, we will prove convergence in probability towards a unique limit. The derivation of the strong fluid limit will prove to be very useful: It allows to calculate the exact stability conditions (policy dependent) (see Theorem V.1 and the numerical Section VII). In addition, the exact characterization of the strong fluid limit provides crucial insights into the performance of the system, such as the growth rates of the number of users over time and monotonicity results with respect to the tie-breaking rule.

Obtaining exact fluid-limit characterizations will require to deal with averaging phenomena: it may happen that one class of users reaches its stationary regime, i.e., is empty in the fluid scaling, before the other classes do. In this case, the drift of the other classes needs to be averaged with the stationary distribution of this class. Hence, a description of the fluid limit will involve averaged drifts as will be defined in (8) (we refer to this as second-vector fields, following [18]).

We focus on the class of policies that induce partially increasing drift vector fields with uniform limits. In order to describe this class of policies we need first to introduce the drift functions and drift vector fields. For the stochastic process $X^f(t)$ associated with a policy $f$, we define the drift function by $\delta^f(x) := (\delta_1^f(x), \ldots, \delta_K^f(x))$, $x \in \mathbb{N}^K$, with

$$\delta_i^f(x) := \mathbb{E}(X_i^f(1) - x_i | X^f(0) = x).$$

(We will drop the superscript $f$ when it is clear that we consider a unique policy.) We say that a vector field $v : \mathbb{N}^K \to \mathbb{R}^K$ has uniform limits [12] if for any $\mathcal{U} \subseteq \{1, \ldots K\}$, there exists a function $v^{\mathcal{U}} : \mathbb{N}^{|\mathcal{U}|} \to \mathbb{R}^K$ (constant when $\mathcal{U} = \emptyset$) such that

$$\lim_{R\to\infty} \sup_{x \in \mathbb{N}^K : |x_{\mathcal{U}^c}| > R} |v(x) - v^{\mathcal{U}}(x_{\mathcal{U}})| = 0,$$

where $x_{\mathcal{U}}$ denotes the restriction of the vector $x$ to indices in the subclass $\mathcal{U}$ and $\mathcal{U}^c$ denotes the complementary set of $\mathcal{U}$. Intuitively, this means that the drift vector-field has limits when we make the number of users of some of the classes go to infinity, and that we can interchange the order of the coordinates when taking these limits. We assume in the following that the drift vector has uniform limits, so that we can define the asymptotic drifts $\delta^{\mathcal{U}} : \mathbb{N}^{|\mathcal{U}|} \to \mathbb{R}^K$ as follows:

$$\delta^{\mathcal{U}}(x_{\mathcal{U}}) := \lim_{x_k \to \infty, \, k \in \mathcal{U}^c} \delta(x). \quad (7)$$

Here, $\mathcal{U}^c$ corresponds to the "saturated" classes for which we let the number of users go to $\infty$. We define the stochastic process $X^{\mathcal{U}}$ as the $\mathcal{U}$-dimensional stochastic process corresponding to the original process seeing an infinite number of users of class $k \in \mathcal{U}^c$ and let $\pi^{\mathcal{U}}$ denote its stationary measure assuming it exists. We define the averaged drift vectors by

$$\bar{\delta}^{\mathcal{U}} = \sum_{x \in \mathbb{N}^{|\mathcal{U}|}} \delta^{\mathcal{U}}(x) \pi^{\mathcal{U}}(x). \quad (8)$$

Finally, following [12] we say that a vector field $v$ is partially increasing if $v_i(x)$ is increasing in $x_j$ for all $j \neq i$. These assumptions, which are crucial to prove the convergence towards the unique strong fluid limit, are verified for many cases of interest, see the next lemma.

**Lemma IV.2.** *A priority index policy or a BR policy with non-state dependent tie-breaking rule (i.e., independently of the numbers of users) induces a partially increasing drift vector field with uniform limits.*

**Proof:** We prove the lemma for BR policies, the other case being similar. When increasing the number of users of one class only, the probability that this class has at least one user in its best possible state is increased. Hence, given that the tie-breaking rule does not depend on the number of users, the probability that this class is served is increased while the probability that a user of another class is served decreases. This implies that the drift vector field is partially increasing.

By the independence of the channel variations, the probability that class $i \in \mathcal{U}^c$ has at least one user in its best state is $1 - (1 - q_{i,N_i})^{x_i}$, where $x_i$ is the number of class-$i$ users. Hence, when the numbers of class-$i$ users, $i \in \mathcal{U}^c$, grows large, the probability of having in each class in $\mathcal{U}^c$ at least one user in its best state (and hence causing a drift $\delta^{\mathcal{U}}(x_{\mathcal{U}})$) converges to 1. Together with the property that the tie-breaking rule does not depend on the number of users, this implies that

$$\delta(x) = \delta^{\mathcal{U}}(x_{\mathcal{U}}) \prod_{i \in \mathcal{U}^c} (1 - (1 - q_{i,N_i}) + o(1/|x|), \quad (9)$$

as $x_k \to \infty, k \in \mathcal{U}^c$, which in turn implies the uniform convergence of $\delta(\cdot)$ to $\delta^{\mathcal{U}}(\cdot)$ for any $\mathcal{U}$. $\quad\square$

We now state the main result of this Section, the description of the strong fluid limit. The proof can be found in Appendix A.

**Theorem IV.3.** *For a given policy $f$ inducing a partially increasing drift vector field with uniform limits, we have*

$$\lim_{r\to\infty} \mathbb{P}(\sup_{0 \leq s \leq t} |Y^{f,r}(s) - y^f(s)| \geq \epsilon) = 0, \quad \text{for all } \epsilon > 0,$$

with $y^f(t)$ a piece-wise linear function that can be described recursively as follows. Let $\mathcal{U}_0 = \emptyset$ and $T_0 = 0$. Then we have,

$$\frac{\mathrm{d}y_k^f(t)}{\mathrm{d}t} = \tilde{\delta}_k^{f,\mathcal{U}_l}, \quad t \in [T_l^f, T_{l+1}^f], \tag{10}$$

$$\text{with } T_{l+1}^f = T_l^f + \min_{k \in \mathcal{U}_l^c, \tilde{\delta}_k^{\mathcal{U}_l} < 0} \frac{y_k^f(T_l^f)}{-\tilde{\delta}_k^{f,\mathcal{U}_l}}, \tag{11}$$

$$\text{and } \mathcal{U}_{l+1} = \mathcal{U}_l \cup \argmin_{k \in \mathcal{U}_l^c, \tilde{\delta}_k^{\mathcal{U}_l} < 0} \frac{y_k^f(T_l^f)}{-\tilde{\delta}_k^{f,\mathcal{U}_l}}, \tag{12}$$

and if there exists no $k \in \mathcal{U}_l^c$ with $\tilde{\delta}_k^{f,\mathcal{U}_l} < 0$, then $T_{l+1}^f = \infty$.

We can now give our definition of the strong fluid limit.

**Definition 7.** *For a given policy $f$ inducing a partially increasing drift vector field with uniform limits, we call the process $y^f(t)$ (as obtained in Theorem IV.3) the strong fluid limit for the policy $f$.*

**Remark 3** (Calculation of the averaged drifts). *Theorem IV.3 characterizes the strong fluid limit as a piece-wise linear function with slopes $\tilde{\delta}^{\mathcal{U}}$. In practice, the calculations of these slopes involve:*

- *deriving the asymptotic drifts (see (7)),*
- *calculating the stationary distributions of $X^{\mathcal{U}}$,*
- *averaging the asymptotic drifts with these stationary distributions (see (8)).*

*For instance, assume $K = 2, N_1 = 2, N_2 = 1$, and a Bernoulli arrival process. Consider the policy that gives priority to the best class-1 user present in the system and otherwise (i.e., when there is no class 1) serves a class-2 user. Then*

$$\delta^{\emptyset} = (\lambda_1 - \mu_{1,N_1}, \lambda_2),$$
$$\delta^{\{1\}}(x_1) = (\lambda_1 - s_1(x_1), \lambda_2 - \mu_{2,N_2}\mathbf{1}_{(x_1=0)}),$$

*with $s_1(x_1) = \mu_{1,N_1}(1-(1-q_{1,N_1})^{x_1})+\mu_{1,N_1-1}(1-q_{1,N_1})^{x_1}$. The process $X^{\{1\}}$ is a 1-dimensional Markov chain with stationary distribution*

$$\pi^{\{1\}}(x_1) = C \prod_{j=1}^{x_1} \frac{\lambda_1(1-s_1(j-1))}{(1-\lambda_1)s_1(j)},$$

*where $C$ is a normalization constant. The average drift can now be computed using (8).*

In the specific case of BR policies with a priority-type tie-breaking rule, we can in fact explicitly derive the strong fluid limit by making use of rate-conservation arguments. This will prove to be very useful to obtain fluid optimality statements.

**Proposition IV.4.** *Consider a BR policy with a priority-type tie-breaking rule. Let us reorder the classes according to the priority ordering. The averaged drift vectors are*

$$\tilde{\delta}^{\emptyset} = (\lambda_1 - \mu_{1,N_1}, \lambda_2, \dots, \lambda_K), \tag{13}$$

*if $T_1 < \infty$, then*

$$\tilde{\delta}^{\{1\}} = (0, \lambda_2 - \mu_{2,N_2}\left(1 - \frac{\lambda_1}{\mu_{1,N_1}}\right), \dots, \lambda_K), \tag{14}$$

*and if $T_{k-1} < \infty$, then $\mathcal{U}_{k-1} = \{1, \dots, k-1\}$ and*

$$\tilde{\delta}^{\mathcal{U}_{k-1}} = (0, \dots, 0, \lambda_k - \mu_{k,N_k}(1 - \sum_{j=1}^{k-1}\frac{\lambda_j}{\mu_{j,N_j}}), \dots, \lambda_K). \tag{15}$$

**Proof:** Using Lemma IV.2, the drift $\delta(\cdot)$ associated to a BR policy is partially increasing with uniform limits, hence the strong fluid limit is given by Theorem IV.3. When $\mathcal{U} = \emptyset$, there are infinitely many users of each class and hence there is always a class-1 user in its best state, which directly implies (13). Note that $T_1 < \infty$ if and only if $\frac{\lambda_1}{\mu_{1,N_1}} < 1$ (see definition of $T_1$ in Theorem IV.3). In this case the process $X^{\{1\}}$ is ergodic. For $T_1 \leq t \leq T_2$, we can simplify the asymptotic drift $\tilde{\delta}^{\{1\}}$ using the specific properties of the policy and rate conservation arguments: let $A_{1,x_1}$ be the event that class 1 is served and there are $x_1$ class-1 users when all the other classes are saturated. With a slight abuse of notation, let us denote by $\pi^{\{1\}}(A_{1,x_1})$ the probability of event $A_{1,x_1}$ under the stationary distribution of $X^{\{1\}}$. Since $X^{\{1\}}$ is ergodic, by rate stability we have the following rate-conservation equation, see (46),

$$\sum_{x_1} \pi^{\{1\}}(A_{1,x_1})\mu_{1,N_1} = \lambda_1,$$

which gives that $\sum_{x_1} \pi^{\{1\}}(A_{1,x_1}^c) = 1 - \frac{\lambda_1}{\mu_{1,N_1}}$. Since $\mathcal{U}_1 = \{1\}$ (so in particular there is still an infinite amount of class-2 users which are exclusively served when there are no class-1 users in their best state) class 2 receives service at rate $\mu_{2,N_2}\sum_{x_1}\pi^{\{1\}}(A_{1,x_1}^c) = \mu_{2,N_2}(1-\frac{\lambda_1}{\mu_{1,N_1}})$ which gives (14).

Consider now the case where $\mathcal{U} := \{1, \dots, k-1\}$ (assuming as before that $\sum_{j=1}^{k-1}\frac{\lambda_j}{\mu_{j,N_j}} < 1$). Let $A_{j,x_\mathcal{U}}$ be the event that class $j \in \mathcal{U}$ is served and there are $x_i$ users of class $i$, $i \in \mathcal{U}$. By rate-conservation arguments, see (46), we obtain

$$\sum_{x_\mathcal{U}} \pi^{\mathcal{U}}(A_{j,x_\mathcal{U}})\mu_{j,N_j} = \lambda_j, \quad j \in \mathcal{U}.$$

Noting that the sets $A_{j,x_\mathcal{U}}$ are disjoints, $\forall\ j \in \mathcal{U}$, this implies that $\sum_{x_\mathcal{U}}\pi^{\mathcal{U}}(\cup_{j\in\mathcal{U}}A_{j,x_\mathcal{U}}) = \sum_{j\in\mathcal{U}}\frac{\lambda_j}{\mu_{j,N_j}}$. Since class $k$ is only served when no class-$i$ users are being served, $i \in \mathcal{U}$, there is a class-$k$ departure with probability $\mu_{k,N_k}(1-\sum_{x_\mathcal{U}}\pi^{\mathcal{U}}(\cup_{j\in\mathcal{U}}A_{j,x_\mathcal{U}}))$. Hence, we obtain Equation (15). $\square$

**Remark 4.** *For all BR policies where the scheduler chooses with probability $\alpha_k^{\mathcal{U}}$ to serve class $k$ when a subset $\mathcal{U}$ of classes has at least one user in its best channel condition, the fluid limit in the interior of the orthant has a drift given by:*

$$\begin{aligned}\delta^{\emptyset} &= (\lambda_1 - \alpha_1^{\{1,\dots,K\}}\mu_{1,N_1}, \lambda_2 - \alpha_2^{\{1,\dots,K\}}\mu_{2,N_2}, \\ &\quad \cdots, \lambda_K - \alpha_K^{\{1,\dots,K\}}\mu_{K,N_K}).\end{aligned} \tag{16}$$

*However, in general we cannot explicitly derive the second-vector fields. An exception is the case of two classes. Then, using the rate-conservation argument as in the previous proposition, we obtain (assuming w.l.o.g. that class 1 empties first)*

$$\tilde{\delta}^{\{1\}} = (0, \lambda_2 - (1 - \frac{\lambda_1}{\mu_{1,N_1}})\mu_{2,N_2}).$$

**Remark 5** (Weak and strong fluid limits). *Though a quite subtle technical point, it is worth emphasizing the conceptual difference between the notion of weak fluid limits and* the notion of strong fluid limit, introduced in Sections IV-A and IV-B, respectively. Note that "weak" versus "strong" refers to accumulation points versus unique limit. The names do however not take into account the mode of convergence.

Weak limits are a powerful tool for stability if one can characterize that they all vanish after a finite amount of time, as will be used in Theorem V.2 for the set of BR policies. However, in general weak limits might not capture the precise asymptotic behavior of the process (see [13]). On the contrary, when having a unique strong fluid limit, the asymptotic behavior of the scaled process is completely described, allowing for example to obtain the policy-dependent stability conditions for a large class of policies (see Theorem V.1).

## V. Stability analysis

The derivation of the weak and strong fluid limits in Section IV allows us to conclude about stochastic stability.

The next Theorem derives the stability conditions for any policy having a partially increasing drift with uniform limits, using the strong fluid limit as obtained in Theorem IV.3.

**Theorem V.1.** *A policy $f$ inducing a partially increasing drift vector field with uniform limits is stable if $T_l^f < \infty$ for all $l$, where $T_l^f$ is given by Theorem IV.3.*

**Proof:** If $T_l^f < \infty$ for all $l$, the strong fluid limit described in Theorem IV.3 is equal to 0 for $t$ large enough, i.e., $Y^{f,r}(t)$ converges in probability to 0 for $t$ large enough. In addition, the random variable $Y_k^{f,r}(t)$ is uniformly integrable. This can be seen by the fact that $Y_k^{f,r}(t)$ can be upper bounded by $x_k(0)$ plus the users that have arrived until time $\lfloor rt \rfloor$ divided by $r$, which is uniform integrable, see [15, Lemma 4.5].

The convergence in probability to 0 and the uniform integrability together imply that $\lim_{r \to \infty} \mathbb{E}(Y_k^{f,r}(t)) = 0$, for $t$ large enough, $\forall\ k$, see [7, Theorem 3.5]. Using an extended Foster-Lyapunov criterion as expressed in [18] or [28, Corollary 9.8], this implies the positive recurrence of $X^f(\cdot)$. □

In the following theorem we state the maximum stability condition, and prove that any BR policy achieves maximum stability. The proof is based on the weak fluid limit characterization as given in Section IV and can be found in Appendix B.

**Theorem V.2.** *The maximum stability condition is*

$$\sum_{k=1}^{K} \frac{\lambda_k}{\mu_{k,N_k}} < 1. \tag{17}$$

*In addition, any BR policy is maximum stable.*

Condition (17) was recognized as the maximal rate stability condition in [1] and as the maximum stability condition under a time-scale separation assumption in [9].

We note that SB, PI and PB are stable under the maximum stability conditions (they belong to the class of BR policies). The intuition behind Theorem V.2 is that asymptotically the system under a BR policy behaves as a classical work-conserving system where class $k$ has departure probability $\mu_{k,N_k}$. On the contrary, other policies, including RB and the $c\mu$-rule, spend (at the fluid scale) a non-negligible fraction of time serving users that are not in their best states, and are therefore not maximum stable. For an example, we refer to Section VII where we numerically obtain the stability conditions for RB and the $c\mu$-rule, making use of Theorem V.1.

**Remark 6** (Overload). *When $\sum_{k=1}^{K} \lambda_k/\mu_{k,N_k} > 1$ the system is said to be in overload. That is, there does not exist any policy that can make the system stable. Theorem IV.3 is however still applicable, providing us with the rates at which the number of users in the different classes grow: given $x(0) = 0$, the growth rate of the number of users over time is given by $X^{f,r}(r)/r = Y^{f,r}(1)$, which in the limit is equal to $\delta^{f,\mathcal{U}}$. This gives us a mean to compare the performance of various policies in overload (see as well Sections VI and VII).*

## VI. Asymptotic fluid optimality

Besides stability, another important performance measure concerns the long-run average holding cost as given in (1). Deriving an optimal policy with respect to this criterion is difficult and the size of the state space makes the problem intractable. For this reason we introduce a related deterministic control problem, which allows us to prove that any BRP policy is asymptotically optimal for the original stochastic system. This emphasizes the important role of the tie-breaking rule in order to achieve efficient performance of the system.

We study the following deterministic fluid control model, which arises from the original stochastic model by only taking into account the mean drifts, i.e.,

$$\min_u \sum_{k=1}^{K} c_k x_k^u(t), \text{ for all } t \geq 0, \quad \text{subject to} \tag{18}$$

$$x_k^u(t) = x_k(0) + \lambda_k t - \sum_{n=1}^{N_k} \mu_{k,n} \int_0^t u_{k,n}(v) \mathrm{d}v, \tag{19}$$

$$x_k^u(t) \geq 0, \ k = 1, \dots, K, \tag{20}$$

$$\sum_{k=1}^{K} \sum_{n=1}^{N_k} u_{k,n}(v) \leq 1, \quad u_{k,n}(v) \geq 0, \ \forall\ k, n, v \geq 0, \tag{21}$$

and the control functions $u_{k,n}(v)$ being integrable. Here $x_k^u(t)$ represents the amount of fluid in class $k$ under control $u(\cdot)$.

We remark that though in general the fluid limit of a policy does depend on the distributions of the random environments (i.e., the $q_{k,n}$'s), these do not appear in the above equations of the fluid control model. This is because the fluid trajectory $x_k(t)$ should be interpreted as a limit of the fluid-scaled process. Hence, when $x_k(t) > 0$ this implies that there are infinitely many class-$k$ users so that with probability 1 there are class-$k$ users in each of the channel state conditions (this being independent of the exact values of the $q_{k,n} > 0$'s).

An optimal control $u^*(\cdot)$ is derived in the following lemma.

**Lemma VI.1.** *Assume $c_1 \mu_{1,N_1} \geq c_2 \mu_{2,N_2} \geq \dots \geq c_K \mu_{K,N_K}$. The fluid control $u^*(\cdot)$ that solves the fluid control problem is as follows. Let $l = \arg \min \{k : x_k(t) > 0\}$. Then*

$$u_{k,N_k}^*(t) = \frac{\lambda_k}{\mu_{k,N_k}}, \text{ for } k < l, \quad u_{l,N_l}^*(t) = 1 - \sum_{i=1}^{l-1} \frac{\lambda_i}{\mu_{i,N_i}},$$

*and $u_{k,n}^*(t) = 0$ otherwise.*

**Proof:** Let us denote $w_k^u(t) = x_k^u(t)/\mu_{k,N_k}$. First, we show that for any feasible control $u(\cdot)$, we have

$$\sum_{k=1}^{j} w_k^{u^*}(t) \leq \sum_{k=1}^{j} w_k^u(t), \text{ for all } t \geq 0, \ j = 1, \ldots, K. \quad (22)$$

If $\sum_{k=1}^{j} w_k^u(t) = 0$, then (22) trivially holds. Now assume $\sum_{k=1}^{j} w_k^{u^*}(t) > 0$. By definition of $u^*(\cdot)$ this implies that $\sum_{k=1}^{j} w_k^{u^*}(s) > 0$, for all $s \in [0,t]$, since once all these classes empty under $u^*(t)$, they will remain empty. Since $u^*(t)$ gives full priority to classes 1 until $j$ over classes $j+1$ until $K$, we have that $\sum_{k=1}^{j} \int_0^t u_{k,N_k}^*(v) dv = t$. Hence, $\sum_{k=1}^{j} \int_0^t u_{k,N_k}^*(v) dv = t \geq \sum_{k=1}^{j} \sum_{n=1}^{N_k} \int_0^t u_{k,n}(v) dv \geq \sum_{k=1}^{j} \sum_{n=1}^{N_k} \int_0^t \frac{\mu_{k,n}}{\mu_{k,N_k}} u_{k,n}(v) dv$, which implies (22), since

$$\sum_{k=1}^{j} w_k^u(t) - \sum_{k=1}^{j} w_k^{u^*}(t)$$
$$= \sum_{k=1}^{j} \int_0^t u_{k,N_k}^*(v) dv - \sum_{k=1}^{j} \sum_{n=1}^{N_k} \int_0^t \frac{\mu_{k,n}}{\mu_{k,N_k}} u_{k,n}(v) dv \geq 0.$$

The minimization term $\sum_{k=1}^{K} c_k n_k^u(t)$ can be written as

$$\sum_{k=1}^{K} c_k \mu_{k,N_k} w_k^u(t) = (c_1 \mu_{1,N_1} - c_2 \mu_{2,N_2}) w_1^u(t)$$
$$+ (c_2 \mu_{2,N_2} - c_3 \mu_{3,N_3})(w_1^u(t) + w_2^u(t)) + \cdots$$
$$+ (c_{K-1} \mu_{K-1,N_{K-1}} - c_K \mu_{K,N_K}) \sum_{k=1}^{K-1} w_k^u(t)$$
$$+ c_K \mu_{K,N_K} \sum_{k=1}^{K} w_k^u(t).$$

Together with (22) and $c_j \mu_{j,N_j} - c_{j+1} \mu_{j+1,N_{j+1}} \geq 0, \forall j$, we obtain that $\sum_{k=1}^{K} c_k \mu_{k,N_k} w_k^u(t)$ is minimized by $u^*(\cdot)$. □

The optimal fluid cost serves as a lower bound for the fluid-scaled cost of the stochastic network, see the lemma below.

**Lemma VI.2.** *For any policy $f$ and for almost all sample paths, we have*

$$\liminf_{r \to \infty} \sum_{k=1}^{K} c_k Y_k^{f,r}(t) \geq \sum_{k=1}^{K} c_k x_k^{u^*}(t), \text{ for all } t \geq 0. \quad (23)$$

**Proof:** Lemma IV.1 states that for almost all sample paths $\omega$ it holds that $\liminf_{r \to \infty} Y_k^{f,r}(t) = y_k^f(t)$, with $y_k^f(t)$ a weak fluid limit for policy $f$ (this follows by considering the subsequence $r_l$ corresponding to the liminf-sequence in Lemma IV.1). Note that a weak fluid limit is an admissible trajectory for the fluid control problem. Hence,

$$\liminf_{r \to \infty} \sum_{k=1}^{K} c_k Y_k^{f,r}(t) = \sum_{k=1}^{K} c_k y_k^f(t) \geq \sum_{k=1}^{K} c_k x_k^{u^*}(t).$$

This concludes the proof. □

Since (23) holds almost surely, it follows by Fatou's lemma that the lower bound holds in probability as well, i.e.,

$$\mathbb{P}(\sum_{k=1}^{K} c_k Y_k^{f,r} - \sum_{k=1}^{K} c_k x_k^{u^*}(t) \geq 0) \to 1, \text{ for all } t \geq 0. \quad (24)$$

We define a policy to be *asymptotically fluid optimal* when the lower bound is obtained in probability, i.e., $\lim_{r \to \infty} \mathbb{P}(|\sum_{k=1}^{K} c_k(Y_k^{f,r}(s) - x^{u^*}(s))| \geq \epsilon) = 0, \forall \epsilon > 0$. The following Theorem characterizes a class of policies that is asymptotically fluid optimal.

**Theorem VI.3.** *Any BRP policy is asymptotically fluid optimal.*

**Proof:** We have $\frac{d x_k^{u^*}(t)}{dt} = \lambda_k - u_{k,N_k}^*(t) \mu_{k,N_k}$, with $u^*(\cdot)$ the optimal control as derived in Lemma VI.1. This drift coincides with the drift of the strong fluid limit $y^{BRP}(\cdot)$, see Proposition IV.4, hence $y^{BRP}(t) = x^{u^*}(t)$. Together with Theorem IV.3, we then obtain that $\lim_{r \to \infty} \sum_{k=1}^{K} c_k Y_k^{BRP,r}(t)$ converges in probability to $\sum_{k=1}^{K} c_k x_k^{u^*}(t)$, i.e., any BR policy is asymptotically fluid optimal. □

It can be checked that the above implies that any BRP policy minimizes $\liminf_{r \to \infty} \mathbb{E}(\int_0^\infty \sum_k c_k Y_k^{f,r}(t) dt)$. Unfortunately, this does not give any performance guarantee in terms of the long-run time-average holding cost as in Equation (1). Numerical experiments reported in Section VII indicate however that BRP policies significantly outperform all other policies.

Note that the optimality results described in this section also apply in overload systems. We have the following corollary for the total growth rate.

**Corollary VI.4.** *Any BRP policy minimizes the growth rate of the total cost, i.e., for all $\epsilon > 0$,*

$$\lim_{r \to \infty} \mathbb{P}\left( \frac{\sum_{k=1}^{K} c_k X_k^{r,f}(r)}{r} - \frac{\sum_{k=1}^{K} c_k X_k^{r,BRP}(r)}{r} \geq -\epsilon \right) = 1.$$

**Proof:** Combining (24) with the asymptotic optimality of a BRP policy, the statement is immediate. □

To the best of our knowledge, the only policy studied in the literature that belongs to BRP, and hence is both maximum stable and asymptotically optimal, is PI. We recall that PI was derived in [4] as the solution of a relaxed optimization problem. SB and PB will as well become asymptotically optimal when the myopic tie-breaking rule would be applied, showing the importance of the tie-breaking rule.

**Remark 7.** *From Theorem VI.3 we conclude that the myopic tie-breaking rule is crucial in order to obtain an asymptotically fluid optimal scheduling policy. In view of equation (2), we note that under this myopic rule, higher priority is given to users with smaller service requirements, $\mathbb{E}(B_k)$. Hence, BRP policies appropriately mix size-based information with achieving opportunistic gains. This is in agreement with the findings of [30] where the authors investigate the tradeoff between prioritizing small users and opportunistic scheduling: They show that if the opportunistic capacity is upper bounded and increases as $1 - a^x$, with $a \in [0,1)$ and $x$ the number*

| Channel state ($n$) | 1 | 2 | 3 | 4 | 5 | 6 | 7 | 8 | 9 | 10 | 11 |
|---|---|---|---|---|---|---|---|---|---|---|---|
| Transmission rate (kb/s) in CDMA | 38.4 | 76.8 | 102.6 | 153.6 | 204.8 | 307.2 | 614.4 | 921.6 | 1228.8 | 1843.2 | 2457.6 |
| Probabilities in CDMA | 0.00 | 0.01 | 0.04 | 0.08 | 0.15 | 0.24 | 0.18 | 0.09 | 0.12 | 0.05 | 0.04 |
| $q_{1,n}$ | 0 | 0 | 0.05 | 0 | 0.23 | 0 | 0.42 | 0 | 0.21 | 0 | 0.09 |
| $q_{2,n}$ | 0 | 0 | 0.15 | 0 | 0.33 | 0 | 0.52 | 0 | 0 | 0 | 0 |
| $\mu_{1,n}$ | 0 | 0 | 0.017 | 0 | 0.033 | 0 | 0.1 | 0 | 0.2 | 0 | 0.4 |
| $\mu_{2,n}$ | 0 | 0 | 0.017 | 0 | 0.033 | 0 | 0.1 | 0 | 0 | 0 | 0 |

TABLE I

TRANSMISSION RATES AND CHANNEL CONDITION PROBABILITIES IN THE CDMA 1XEV-DO WIRELESS NETWORK, AS REPORTED IN [5].

of users, then a significant improvement of performance can be achieved by exploiting information on the service time requirement. In our model the capacity has this behavior, see for instance Equation (9), and as will be observed in the numerical results, exploiting size-based information indeed allows to obtain significant improvements.

## VII. NUMERICAL EXPERIMENTS

We consider a CDMA 1xEV-DO system with two classes of users ($K = 2$). Time is slotted, with the length of one slot being $t_c = 1.67ms$. In each time slot, one new class-$k$ user arrives with probability $\lambda_k$. We choose 10.257 kb as the expected service requirement of both a class-1 and class-2 user. Associated to the state of the channel, we have transmission rates (kb/s), see Table I (taken from [5]). We assume that class-1 users have five possible transmission rates while class-2 users have three. The corresponding probabilities ($q_{k,n}$) are given in Table I. In addition, applying equation (2) we calculate the departure probabilities ($\mu_{k,n}$). We fix $\lambda_2 = 0.05$, so $\lambda_2/\mu_{2,N_2} = 0.5$. We set $c_1 = c_2 = 1$, so that we are interested in minimizing the expected total number of users in the system, see Equation (1). In addition, we compare the performance of the policies SB, RB, PI, PB and the $c\mu$ rule, which were introduced in Section III.

Before presenting the numerical results we first summarize the main conclusions that we will make in this section:

- We prove that not all policies obtain maximum stability. More precisely, we calculate the stability conditions under RB and the $c\mu$-rule and observe that these are much more stringent than the stability condition for BR policies (e.g. SB, PB and PI).
- The drifts of the fluid limit, $\delta^{\mathcal{U}}$, which are calculated numerically (and in some cases theoretically), provide very important insights on the performance. In particular, insightful monotonicity results in the tie-breaking rule are obtained with respect to the performance of the system. In addition, the drift analysis allows to show that PI outperforms all other policies in terms of the growth rates.
- Our simulations illustrate that the tie-breaking rule has a very big impact on the performance of the system and we find that combining opportunistic scheduling with the myopic tie-breaking rule gives optimal mean performance, as was also suggested by the asymptotic fluid optimality result of BRP policies in Section VI.

*a) Fluid limit:* We first illustrate how the scaled process converges to the fluid limit. We take $r = 10000$, $Y^r(0) = X(0)/r = (1, 1)$ and plot the scaled processes $Y_1^r(t)$, $Y_2^r(t)$, and $Y_1^r(t) + Y_2^r(t)$ for different policies, see Figure 2. In this simulation we set $\lambda_1 = 0.14$, so $\lambda_1/\mu_{1,N_1} = 0.35$.

We describe the fluid limit $y^f(t)$ as defined in Theorem IV.3. When both classes are saturated, i.e., $\mathcal{U} = \emptyset$, the drift is

$$\delta^{f,\emptyset} = (\lambda_1 - \alpha^f \mu_{1,N_1}, \lambda_2 - (1 - \alpha^f)\mu_{2,N_2}), \quad (25)$$

see Remark 4. Here $\alpha^f$ is a random tie-breaking rule, i.e., in case of a tie, $\alpha^f$ is the probability that class 1 is favoured over class 2. For our set of parameters, the best class-1 user under the $c\mu$-rule and RB is always preferred over the best class-2 user, i.e., there occur no ties, hence one can set $\alpha^f = 1$ in (25) for $f = c\mu, RB$. For PI, SB and PB we do have ties, and we set $\alpha^{PI} = 1$ (since PI applies the myopic tie-breaking rule) and $\alpha^{SB} = \alpha^{PB} = 1/2$ (since SB and PB apply a random tie-breaking rule). In Table II we present the so-obtained values for $\delta^{f,\emptyset}$. From the drifts it is clear that under all policies class 1 empties before class 2. The moment that this happens, $T_1^f$, can be derived from Theorem IV.3 and satisfies $T_1^{PI} = T_1^{c\mu} = T_1^{RB} < T_1^{SB} = T_1^{PB}$, see also Figure 2 a).

For $T_1^f < t \le T_2^f$, the drift of class 1 is 0, whereas the drift of class 2 is going to depend on the policy. From Proposition IV.4 we have that for all BR policies (e.g. PI, SB and PB) $\delta^{f,\{1\}} = (0, \lambda_2 - \mu_{2,N_2}(1 - \lambda_1/\mu_{1,N_1}))$. For the $c\mu$ rule and RB we calculate the drift numerically using Remark 3. In particular we observe that these drifts are positive for the latter two policies, which implies instability of the system, as can be seen in Figure 2 b). We observe that for $t \le T_1^f$ the number of class-2 users increases under policies PI, $c\mu$ and RB, while for SB and PB, the drift of class-2 users is negative.

A direct consequence of the drift function is that SB, PB, and PI (in fact all BR policies) empty the system at the same time (under the maximum stability condition), i.e., $T_2^f$ is the same (this can be derived directly from Equation (45) for example). However, the performance of a policy will depend on the order in which classes are served. In the fluid limit, this is fully determined by the choice of the tie-breaking rule. Note that, as can be seen from Figure 2 c), PI, (and hence any BR policy with the myopic tie-breaking rule) minimizes the total number of users at any moment in time.

*b) Stability region:* We now vary the value of $\lambda_1$ from 0.004 to 0.196, and as a consequence we have that $\rho := \lambda_1/\mu_{1,N_1} + \lambda_2/\mu_{2,N_2}$ varies from 0.51 to 0.99. The policies

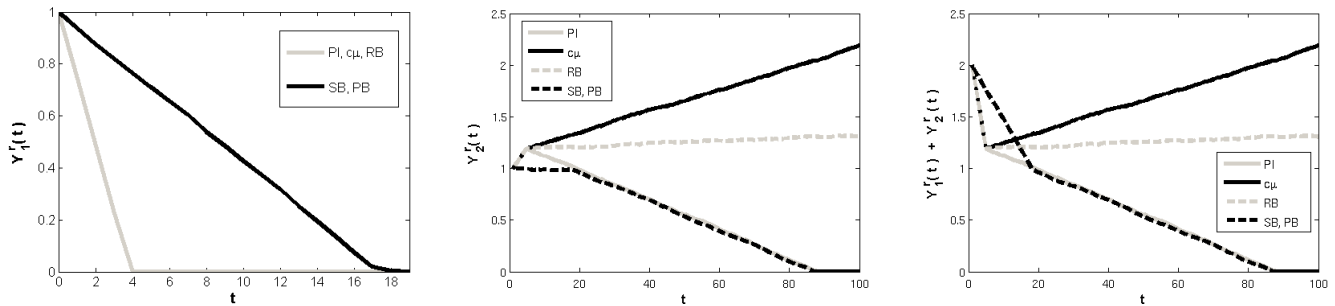

Fig. 2. (a) Scaled number of class-1 users, (b) Scaled number of class-2 users, (c) Scaled total number of users

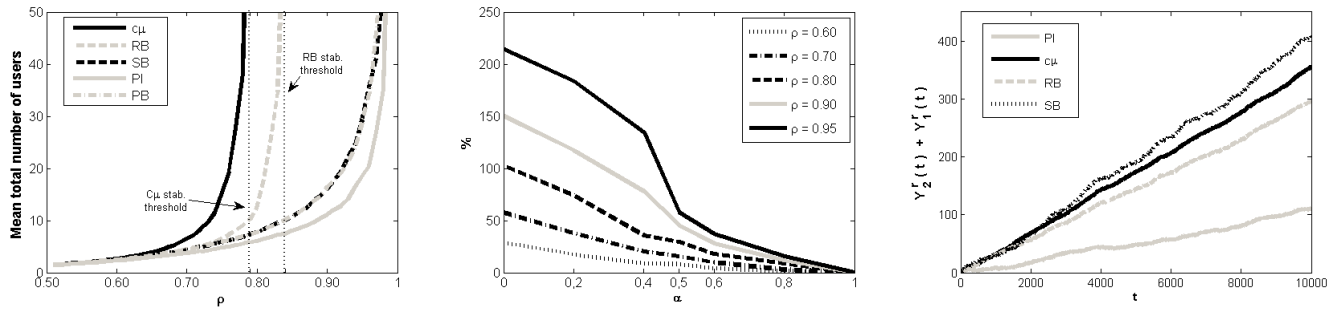

Fig. 3. (a) Mean number of users and stability thresholds, (b) PI under different tie-breaking rules: relative degradation (in %) over PI with $\alpha = 1$, (c) Scaled total number of users in overload, $\rho = 1.1$.

| | $\delta^{f,\emptyset}$ | | $\delta^{f,\{1\}}$ | |
|---|---|---|---|---|
| f | Class 1 | Class 2 | Class 1 | Class 2 |
| **PI** | -0.26 | 0.05 | 0 | -0.015 |
| $c\mu$-rule | -0.26 | 0.05 | 0 | 0.0096 |
| PB/SB | -0.06 | 0 | 0 | -0.015 |
| RB | -0.26 | 0.05 | 0 | 0.0004 |

TABLE II
DRIFT OF THE FLUID LIMIT.

| | $\delta^{f,\emptyset}$ | | $\delta^{f,\{1\}}$ | |
|---|---|---|---|---|
| f | Class 1 | Class 2 | Class 1 | Class 2 |
| PI | -0.16 | 0.05 | 0 | 0.01 |
| $c\mu$-rule | -0.16 | 0.05 | 0 | 0.036 |
| PB/SB | 0.04 | 0 | - | - |
| RB | -0.16 | 0.05 | 0 | 0.029 |

TABLE III
DRIFT OF THE FLUID LIMIT IN OVERLOAD SYSTEM.

PI, PB and SB belong to the BR policies, and are hence stable when $\rho < 1$. For the $c\mu$-rule and RB the stability condition can be calculated (numerically) by setting $\delta_2^{f,\{1\}}$ equal to zero and following the steps described in Remark 3. In particular, the $c\mu$-rule is stable if and only if $\rho < 0.79$ and RB is stable if and only if $\rho < 0.84$. In Figure 3 a) we plot the mean number of users for different values of $\rho$ and we observe that the mean number of users (and hence the mean delay) for these policies grows to infinity as the load approaches the critical value.

*c) Impact of Tie-Breaking rule:* We study the impact of the tie-breaking rule on the performance of the system.

In order to investigate this issue in more depth, we simulate PI under different random tie-breaking rules, i.e., we let the probability $\alpha$ vary from 0 until 1 (Recall that the parameter $\alpha$ is the probability, in case of a tie, that class 1 is favoured over class 2). We emphasize that PI as defined in [4] uses by default the myopic tie-breaking rule, i.e., $\alpha^{PI} = 1$. In Figure 3 b) we plot the relative degradation (in terms of the mean number of users) over PI as we vary $\alpha$. We observe that the degradation in the mean performance is decreasing as $\alpha$ increases. In the fluid-scaling system a similar observation can be made: from (25) and Remark 4 it follows that the drift of the total number of users of the strong fluid limit (as obtained in Theorem IV.3) is decreasing (or constant) in $\alpha$, for all time $t$.

The results show that the myopic tie-breaking rule, which was proven to be asymptotically optimal in the fluid limit (see Theorem VI.3), is in practice indeed optimal when minimizing the mean number of users. In addition, the relative degradation of the tie-breaking rule with $\alpha = 1/2$ (compared to the myopic tie-breaking rule $\alpha^{PI} = 1$) can be very large. For example, for $\rho = 0.8$ the degradation is 29% and for $\rho = 0.9$ it is 45%.

*d) Overload:* In Figure 3 c) we plot a trajectory of the total scaled number of users $Y_1^r(t) + Y_2^r(t)$ (for $r = 100$) when $\lambda_1 = 0.240$ and $\rho = 1.1$, so all policies are unstable. In Table III we give the values for the drifts (growth rates) of the fluid limit $y^f(t)$ in this overload setting. In this example, the worst performance is under SB and the best performance is for PI. This in contrary to the stable regime where SB is maximum stable with a performance strictly better than the $c\mu$-rule and RB, see Figure 3 a). This implies that the performance of this policy can differ very much between stable and overload regimes. In addition, we note that the total growth rate of the system is always minimized under any BRP policy, that is a best-rate policy with the myopic tie-breaking rule, as is proved in Corollary VI.4.

## VIII. Conclusion

We have characterized the classes of policies that are maximum stable and asymptotically optimal in a system with random environment. An important conclusion, validated by numerical experiments, is that the tie-breaking rule has a tremendous impact on the performance. Our analysis also shows that simple priority-index policies like PI or SB with a $c\mu$ tie-breaking rule, are stable and asymptotically optimal. While in this model we assumed geometric service requirements, we do believe that direct extensions of all our results exist for phase-type distributed service requirements. In particular, we expect that an optimal tie-breaking rule will be of a simple priority-index type.

**Acknowledgements** We are grateful to S.C. Borst (TU/e and Bell-Labs) and P. Jacko (BCAM) for many fruitful discussions.

## Appendix A: Proof of Theorem IV.3

We first derive a property on the drift vector.

**Lemma VIII.1.** *Suppose $\delta^f(\cdot)$ has uniform limits. Then for any $\mathcal{U}$ and any compact set $\mathcal{C} \subset \mathbb{R}_{++}^{|\mathcal{U}^c|}$ there exists an $L > 0$ and $\epsilon_r$ with $\lim_{r\to\infty} \epsilon_r = 0$ such that*

$$|\delta^f(x) - \delta^{\mathcal{U},f}(x_{\mathcal{U}})| \leq \epsilon_r + L|\frac{x_{\mathcal{U}^c}}{r} - y|,$$

$$\forall x \in \mathbb{R}_+^K, \ x_{\mathcal{U}^c} \in \mathbb{R}_{++}^{|\mathcal{U}^c|}, \ y \in \mathcal{C} \subset \mathbb{R}_{++}^{|\mathcal{U}^c|}. \tag{26}$$

**Proof:** Consider $\epsilon > 0$, and $y \in \mathcal{C}$ such that $y_i > \epsilon$, $\forall i \in \mathcal{U}^c$. Define $\mathcal{B}_{ry,r\epsilon} \subset \mathbb{R}_{++}^{|\mathcal{U}^c|}$ the ball of center $ry$ and radius $r\epsilon$. Assume that the drift function $\delta(\cdot)$ has uniform limits. This implies that for all $x_{\mathcal{U}}$ the sequence

$$\epsilon_r = \sup_{y\in\mathcal{C}} \sup_{x_{\mathcal{U}^c}\in\mathcal{B}_{ry,r\epsilon}} |\delta^f(x) - \delta^{\mathcal{U},f}(x_{\mathcal{U}})|,$$

is converging to 0, when $r \to \infty$. Hence, remarking that $|\frac{x_{\mathcal{U}^c}}{r} - y| \geq \epsilon$ when $x_{\mathcal{U}^c} \notin \mathcal{B}_{ry,r\epsilon}$, we obtain that

$$
\begin{aligned}
|\delta^f(x) - \delta^{\mathcal{U},f}(x_{\mathcal{U}})| &\leq \sup_{x_{\mathcal{U}^c} \in \mathcal{B}_{ry,r\epsilon}} |\delta^f(x) - \delta^{\mathcal{U},f}(x_{\mathcal{U}})| + \\
&\quad + \sup_{x_{\mathcal{U}^c} \notin \mathcal{B}_{ry,r\epsilon}} |\delta^f(x) - \delta^{\mathcal{U},f}(x_{\mathcal{U}})| \\
&\leq \epsilon_r + L|\frac{x_{\mathcal{U}^c}}{r} - y|,
\end{aligned}
$$

for $L := \frac{2D}{\epsilon} \geq \frac{\sup_{x_{\mathcal{U}^c} \notin \mathcal{B}_{ry,r\epsilon}} |\delta^f(x) - \delta^{\mathcal{U},f}(x_{\mathcal{U}})|}{\epsilon}$, where $D := \max_{k,n}(\lambda_k, \mu_{k,n})$. $\square$

In what follows, we will use the martingale decomposition of the process (see [35]), i.e., we define $M(\cdot)$ such that

$$
X(t) = X(0) + \sum_{s=0}^{t-1} \delta(X(s)) + M(t), \tag{27}
$$

where $M$ is a martingale, and $|M(0)| = 0$. The following lemma gives a useful property for the martingale $M(t)$.

**Lemma VIII.2.** *It holds that* $\mathbb{E}\left(|M_k(t)|^2\right) \leq C't$, $\forall k = 1 \ldots, K$, *for some constant* $C' > 0$.

**Proof:** Using that the drift is bounded and the arrival increments have a finite variance, it immediately follows from (27) that $\mathbb{E}\left(|M_k(t)|^2\right) < \infty$, $\forall t$. Hence,

$$
\begin{aligned}
&\mathbb{E}\left(|M_k(t)|^2\right) \tag{28} \\
&\leq \mathbb{E}\left(|M_k(0)|^2\right) + \sum_{s=1}^{t} \mathbb{E}\left(|(M_k(s) - M_k(s-1))|^2\right),
\end{aligned}
$$

see [35, Chapter 12.1]. From (27), we have that for all $s$,

$$
\begin{aligned}
&|M_k(t) - M_k(t-1)| \\
&\leq |X_k(t) - X_k(t-1)| + |\delta(X_i(t-1))|.
\end{aligned}
$$

Since the drift is bounded, and the arrival process has finite mean and variance, this implies that for $C' > 0$ a constant

$$
\mathbb{E}\left(|M_k(t) - M_k(t-1)|^2\right) \leq C'.
$$

Thus, the statement follows from (28). $\square$

We prove Theorem IV.3 for two classes of users, the general case being notationally much more cumbersome but mathematically equivalent. The trajectory of the fluid limit is defined piece-wise. We refer as the *first part of the fluid limit* to the case $t < T_1$, and as *second part of the fluid limit* to the case $T_1 \leq T_2$, and we shall prove the convergence separately for the two parts. The basic techniques used to prove the convergence of the first part of the trajectory are fairly standard and we proceed using the tools of [20]. Let us however emphasize that the proof relies crucially on the extended Lipschitz condition (26). Verifying this condition might be difficult in practice. In Lemma VIII.1, combined with Lemma IV.2, we however showed that the conditions are satisfied for various scheduling policies of interest. Proving the convergence of the second part of the trajectory is much more subtle as it involves an averaging phenomena: one of the classes reaches a stationary regime while the other class keeps being macroscopically big. We must therefore rely on stochastic comparisons and on the ergodic theorem for Markov chains. Such phenomena have been observed and analyzed in [18] for homogeneous random walks in the positive orthant.

*First part of the trajectory:* Assume strictly positive initial conditions for both classes, i.e., $x_1(0) > 0$ and $x_2(0) > 0$. At time $t = 0$ all classes are saturated, hence in the first part of the trajectory, $t < T_1$, we need to consider $\mathcal{U} = \mathcal{U}_0 = \emptyset$. Consider the deterministic function $y(\cdot)$ as defined in the theorem, $y(t) = x(0) + \delta^{\emptyset} t$, and we define an error function

$$
e^r(t) := \mathbb{E}\left(\sup_{0 \leq s \leq t} |Y^r(s) - y(s)|\right).
$$

We obtain from (27) that

$$
|Y^r(t) - y(t)| \leq \frac{1}{r} \sum_{s=0}^{\lfloor rt \rfloor - 1} \left|\delta(X^r(s)) - \delta^{\emptyset}\right| + \left|\frac{M^r(\lfloor rt \rfloor)}{r}\right|. \tag{29}
$$

We first control the martingale. Using successively Cauchy-Schwartz, Doob's inequality and Lemma VIII.2, we get

$$
\begin{aligned}
\mathbb{E}\left(\sup_{s \leq t}\left|\frac{M^r(\lfloor rs \rfloor)}{r}\right|\right)^2 &\leq \mathbb{E}\left(\left(\sup_{s \leq t}\left|\frac{M^r(\lfloor rs \rfloor)}{r}\right|\right)^2\right) \\
&\leq 4\mathbb{E}\left(\left|\frac{M^r(\lfloor rt \rfloor)}{r}\right|^2\right) \leq \frac{C't}{r}, \quad \text{for a } C' > 0. \tag{30}
\end{aligned}
$$

We now control the drift part in Equation (29). Since the drift has uniform limits, we obtain from Lemma VIII.1 that

$$
|\delta(X^r(s)) - \delta^{\emptyset}| \leq \epsilon_r + L|\frac{X^r(s)}{r} - y(\frac{s}{r})|, \ \forall \ s \leq rt.
$$

Hence, bounding the Riemann sum for piece-wise constant functions by an integral, there exists a $\kappa > 0$ such that

$$
\begin{aligned}
\frac{1}{r}\sum_{s=0}^{\lfloor rt \rfloor - 1}\left|\delta(X^r(s)) - \delta^{\emptyset}\right| &\leq \epsilon_r t + \kappa \int_0^t |\frac{X^r(rs)}{r} - y(s)|\mathrm{d}s \\
&\leq \epsilon_r t + \kappa \int_0^t \sup_{0 \leq \tilde{s} \leq s}\left|\frac{X^r(r\tilde{s})}{r} - y(\tilde{s})\right|\mathrm{d}s. \tag{31}
\end{aligned}
$$

Taking expectations on both sides we obtain

$$
\mathbb{E}\left(\frac{1}{r}\sum_{s=0}^{\lfloor rt \rfloor - 1}\left|\delta(X^r(s)) - \delta^{\emptyset}\right|\right) \leq \epsilon_r t + \kappa\left(\int_0^t e^r(s)\mathrm{d}s\right). \tag{32}
$$

Together with (29), (30) and (32), we deduce that

$$
\begin{aligned}
e^r(t) &\leq \mathbb{E}\left(\frac{1}{r}\sum_{s=0}^{\lfloor rt \rfloor - 1}\left|\delta(X^r(s)) - \delta^{\emptyset}\right|\right) + \mathbb{E}\left(\sup_{s \leq t}\left|\frac{M^r(\lfloor rs \rfloor)}{r}\right|\right) \\
&\leq \epsilon_r t + \kappa\left(\int_0^t e^r(s)\mathrm{d}s\right) + \sqrt{\frac{C't}{r}}.
\end{aligned}
$$

Hence, for any $\epsilon > 0$ there exists $r_0$ such that, for a $C'' > 0$,

$$
e^r(t) \leq C''\left(\epsilon t + \int_0^t e^r(s)ds\right), \ \text{ for } r \geq r_0.
$$

Using Gronwall's lemma [27], we obtain that for any $r \geq r_0$,

$$
e^r(t) \leq C''\epsilon t\left(1 + e^t\int_0^t e^{-s}ds\right) \leq C''\epsilon t\left(1 + te^t\right),
$$

for all $t < T_1$. Hence, $\lim_{r\to\infty} e^r(t) = 0$ for $t < T_1$. Since $e^r(t) = \mathbb{E}(\sup_{0\le s\le t}|Y^r(s) - y(s)|) \ge \mathbb{P}(\sup_{0\le s\le t}|Y^r(s) - y(s)| > \epsilon') \cdot \epsilon'$, we obtain $\lim_{r\to\infty} \mathbb{P}(\sup_{0\le s\le t}|Y^r(s) - y(s)| > \epsilon') = 0$, for $t < T_1$.

In case $T_1 = \infty$, the Theorem IV.3 is now proved $\forall t > 0$. In the remainder of the proof we therefore assume that $T_1 < \infty$.

*Second part of the trajectory (one stationary class):* We decomposed the trajectory of $Y^r$ into a first part ($t < T_1$) and a second part ($t > T_1$). By the Markov property, we can study the second part of the trajectory supposing that $Y^r(T_1) = (x_1, y_2(T_1)r)$. Since $T_1 < \infty$, the process $X_1^{\{1\}}$ is ergodic. Using the monotonicity of the drift, we have that $X_1^r(t) \le_{st} X_1^{\{1\}}(t)$ for $t > T_1$ (this can be observed by coupling the two processes and noting that in the case of an infinite number of class-2 users, the service rate for class 1 is the lowest possible since there will always be a class-2 user in its best state). This implies that the family $(X_1^r(t))_r$ is tight and that $\frac{X_1^r(t)}{r} \to 0$, as $r \to \infty$, for $t > T_1$. At time $T_1$ class 2 is saturated, i.e., $\mathcal{U}_1 = \{1\}$. In what follows we assume that $T_1 \le t < T_2$. Consider the deterministic function $y_2(\cdot)$ as defined in the theorem, $y_2(t) = y_2(T_1) + \tilde{\delta}_2^{\{1\}}(t - T_1)$, $T_1 \le t < T_2$, and define the error function

$$E_2^r(t) := \sup_{T_1 \le s \le t} |Y_2^r(s) - y_2(s)|.$$

From (27) we obtain

$$Y_2^r(t) = y_2(T_1) + \frac{1}{r}\sum_{s=\lfloor rT_1\rfloor}^{\lfloor rt\rfloor -1} \delta_2(X^r(s)) + \frac{M_2^r(\lfloor rt\rfloor)}{r}. \quad (33)$$

Using the triangular inequality we obtain,

$$E_2^r(t) \le \frac{1}{r}\sum_{s=\lfloor rT_1\rfloor}^{\lfloor rt\rfloor -1} \left| \delta_2(X^r(s)) - \delta_2^{\{1\}}(X_1^r(s)) \right| \quad (34)$$

$$+ \frac{1}{r}\sum_{s=\lfloor rT_1\rfloor}^{\lfloor rt\rfloor -1} \left| \delta_2^{\{1\}}(X_1^r(s)) - \tilde{\delta}_2^{\{1\}} \right| + \sup_{0\le s\le t} \left| \frac{M_2^r(\lfloor rs\rfloor)}{r} \right|.$$

From Lemma VIII.1 we obtain that there exists $\epsilon_r$ with $\lim_{r\to\infty} \epsilon_r = 0$, such that $\forall s = \lfloor rT_1\rfloor, \ldots, \lfloor rt\rfloor$.

$$|\delta_2(X^r(s)) - \delta_2^{\{1\}}(X_1^r(s))| \le \epsilon_r + L|\frac{X_2^r(s)}{r} - y_2(\frac{s}{r})|. \quad (35)$$

We will now derive a bound on $|\delta_2^{\{1\}}(X_1^r(s)) - \tilde{\delta}_2^{\{1\}}|$. In order to do that, we define a one-dimensional process $\underline{X}_1^r$ which has the same dynamics as our initial process but for which we fix the class-2 users to be equal to $r(y_2(T_1) - \theta)$, with $0 < \theta < y_2(T_1)$. Let $\underline{\pi}^{\theta,r}$ be the stationary distribution of $\underline{X}_1^r$. Adapting Lemma 6 in [12] to our discrete time setting, we obtain that

$$\lim_{r\to\infty} \underline{\pi}^{\theta,r}(x) = \pi^{\{1\}}(x), \forall x. \quad (36)$$

We define $T^r = \inf\{t : X_2^r(rt) < r(y_2(T_1) - \theta)\}$. Using coupling arguments, we see that $T^r \ge T_1$ almost surely, while for $T_1 < t \le T^r$, $\underline{X}_1^r(t) \le_{st} X_1^r(t) \le_{st} X_1^{\{1\}}(t)$. Together with the fact that $\delta_2^{\{1\}}(\cdot)$ is increasing (using the monotonicity

assumption) we obtain that

$$\frac{1}{r}\sum_{s=\lfloor rT_1\rfloor}^{\lfloor rt\rfloor -1} \delta_2^{\{1\}}(\underline{X}_1^r(s)) \le \frac{1}{r}\sum_{s=\lfloor rT_1\rfloor}^{\lfloor rt\rfloor -1} \delta_2^{\{1\}}(X_1^r(s))$$

$$\le \frac{1}{r}\sum_{s=\lfloor rT_1\rfloor}^{\lfloor rt\rfloor -1} \delta_2^{\{1\}}(X_1^{\{1\}}(s)) \text{ for } t \le T^r. \quad (37)$$

Since $\delta_2^{\{1\}}$ is uniformly upper-bounded, we have that

$$\lim_{r\to\infty}\sum_x \underline{\pi}^{\theta,r}(x)\delta_2^{\{1\}}(x) = \sum_x \lim_{r\to\infty} \underline{\pi}^{\theta,r}(x)\delta_2^{\{1\}}(x)$$

$$= \sum_x \pi^{\{1\}}(x)\delta_2^{\{1\}}(x) = \tilde{\delta}_2^{\{1\}}.$$

Hence, we obtain that for all $\epsilon > 0$ there exists $r_0$ such that $\sum_{x\ge 0} \underline{\pi}^{\theta,r_0}(x)\delta_2^{\{1\}}(x) \ge \tilde{\delta}_2^{\{1\}} - \epsilon$, and for all $r \ge r_0$, and applying the ergodic theorem for Markov chains we get

$$\lim_{r\to\infty} \frac{1}{r}\sum_{s=\lfloor rT_1\rfloor}^{\lfloor rt\rfloor -1} \delta_2^{\{1\}}(X_1^{\{1\}}(s)) = (t-T_1)\tilde{\delta}_2^{\{1\}}, \text{ a.s. and}$$

$$\lim_{r\to\infty} \frac{1}{r}\sum_{s=\lfloor rT_1\rfloor}^{\lfloor rt\rfloor -1} \delta_2^{\{1\}}(\underline{X}_1^r(s)) \ge (t-T_1)(\tilde{\delta}_2^{\{1\}} - \epsilon), \text{ a.s..}$$

Together with (37) we now obtain that there exists an $r_0$ such that for all $r \ge r_0$

$$|\frac{1}{r}\sum_{s=\lfloor rT_1\rfloor}^{\lfloor rt\rfloor -1} \delta_2^{\{1\}}(X_1^r(s)) - \tilde{\delta}_2^{\{1\}}| \le \epsilon, \text{a.s., for } t \le T^r. \quad (38)$$

Let us now condition on the set of events $\Omega_1^r(\tilde{\epsilon}) = \{\omega : \sup_{s\le t} \frac{|M_2^r(\lfloor rs\rfloor)|}{r} \le \tilde{\epsilon}\}$. From (34), (35) and (38), and using an inequality similar to (31), we obtain that for any sample path in $\Omega_1^r(\tilde{\epsilon})$ (with $\tilde{\epsilon}$ small enough) it holds that

$$E_2^r(t) \le C''(\epsilon t + \int_0^t E_2^r(s)\mathrm{d}s), \quad \text{for all } t \le T^r, \quad (39)$$

with $r$ large enough, and $C'' > 0$. Using Gronwall's lemma [27], equation (39) implies that $E_2^r(t) \le C''\epsilon t(1+te^t)$. Hence, we can conclude that for a given $\epsilon > 0$, there is an $\tilde{\epsilon} > 0$ such that for each sample path in $\Omega_1^r(\tilde{\epsilon})$ it holds that

$$E_2^r(t) \le \epsilon, \text{ for all } t \le T^r, \text{ and } r \text{ large enough.} \quad (40)$$

We define $T^\infty := T_1 - \frac{\theta}{\tilde{\delta}_2^{\{1\}}}$ if $\tilde{\delta}_2^{\{1\}} < 0$, and $T^\infty = \infty$ otherwise. Assume $\omega \in \Omega_1^r(\tilde{\epsilon})$. If $T^r < T^\infty < \infty$, then

$$T^\infty - T^r = \frac{E_2^r(T^r)}{|\tilde{\delta}_2^{\{1\}}|} \le \frac{\epsilon}{|\tilde{\delta}_2^{\{1\}}|}, \text{ so } T^r \ge T^\infty - \frac{\epsilon}{|\tilde{\delta}_2^{\{1\}}|},$$

see also Figure 4. When $T^\infty = \infty$, it is easy to check that $T^r = \infty$ for $\epsilon$ small enough: Consider an $\epsilon < \theta$. If $T^r < \infty$, then $|Y_2^r(T^r) - y_2(T^r)| > \theta$. However, since $\omega \in \Omega_1^r(\tilde{\epsilon})$ we also have $|Y_2^r(T^r) - y_2(T^r)| < \epsilon$, which gives a contradiction with the above. See also Figure 5.

Therefore, given $t$ such that $t < T^\infty$, there exists an $\epsilon$ small enough such that $t < T^r$ for all $r$ large enough. Hence, from (40) we obtain for $t < T^\infty$,

$$\mathbb{P}\left(\sup_{T_1\le s\le t} |Y_2^r(s) - y_2(s)| \ge \epsilon\right)$$

$$= \mathbb{P}\left(E_2^r(t) \ge \epsilon\right) \le \mathbb{P}\left(\Omega/\Omega_1^r(\tilde{\epsilon})\right). \quad (41)$$

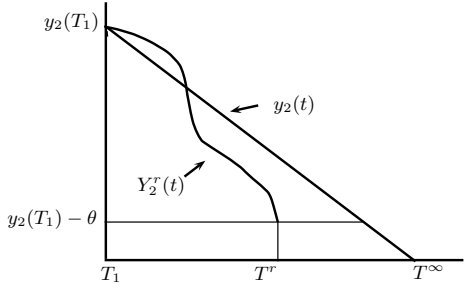

Fig. 4. Case $T^\infty < \infty$.

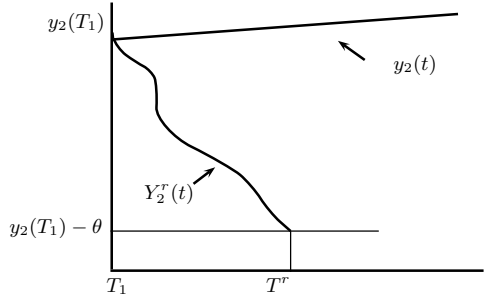

Fig. 5. Case $T^\infty = \infty$.

From (30), we obtain $\lim_{r \to \infty} \mathbb{E}\left(\sup_{s \le t} \left| \frac{M^r(\lfloor rs \rfloor)}{r} \right|\right) = 0$. Since $\mathbb{E}\left(\sup_{s \le t} \left| \frac{M^r(\lfloor rs \rfloor)}{r} \right|\right) \ge \mathbb{P}(\sup_{s \le t} \left| \frac{M^r(\lfloor rs \rfloor)}{r} \right| \ge \tilde{\epsilon}) \cdot \tilde{\epsilon}$, we obtain that $\mathbb{P}\left(\Omega/\Omega_1^r(\tilde{\epsilon})\right) = \mathbb{P}(\sup_{s \le t} \left| \frac{M^r(\lfloor rs \rfloor)}{r} \right| \ge \tilde{\epsilon}) \le \frac{1}{\tilde{\epsilon}} \mathbb{E}\left(\sup_{s \le t} \left| \frac{M^r(\lfloor rs \rfloor)}{r} \right|\right) \to 0$, as $r \to \infty$. Together with (41), this implies that for $t < T^\infty$,

$$\lim_{r \to \infty} \mathbb{P}\left(\sup_{T_1 \le s \le t} |Y_2^r(s) - y_2(s)| \ge \epsilon\right) = 0.$$

Finally, letting $\theta \uparrow y_2(T_1)$, and hence $T^\infty \uparrow T_1 + \frac{y_2(t)}{\delta_2^{(1)}} = T_2$, we obtain that $Y_2^r(\cdot)$ converges in probability on compact sets to $y_2(\cdot)$ in the set $[T_1, T_2)$. □

## APPENDIX B: PROOF OF THEOREM V.2

If condition (17) is not satisfied, the system is not rate stable, see Lemma VIII.3 below, which precludes stability. Hence, (17) is a necessary condition for stability. We are now left with proving that BR policies are stable under condition (17). In order to do that, we will take the following steps: first, we prove that, under any BR policy, the set of sample paths where service is given to a user that is not in its best state is asymptotically almost surely empty. Then we prove that the fluid limit will be equal to zero almost surely for some time $T$ large enough, if condition (17) holds. This will then allow to conclude the proof.

In the remainder of the proof we focus on BR policies and assume condition (17) to be satisfied. Let us define

$$T := \frac{\sum_{k=1}^K \frac{y_k^I(0)}{\mu_{k,N_k}}}{\left(1 - \sum_{k=1}^K \frac{\lambda_k}{\mu_{k,N_k}}\right)} < \infty. \tag{42}$$

We define the random variable $T_\epsilon^r = \inf\{t : \sum_{k=1}^K \frac{Y_k^r(t)}{\mu_{k,N_k}} \le \epsilon\}$ for $0 < \epsilon < 1$ and let $\overline{T}_\epsilon^r = \min\{T_\epsilon^r, T\}$. Consider the event $A_r$ which occurs if, during the time interval $[0, \lfloor r\overline{T}_\epsilon^r \rfloor]$, there is at least one user which has been served while not being in its best possible state. Since any BR policy gives absolute priority to users which are in their best possible state, it follows that

$$\mathbb{P}(A_r) = 1 - \prod_{s=0}^{\lfloor r\overline{T}_\epsilon^r \rfloor} [1 - \prod_{k=1}^K (1 - q_{k,N_k})^{rY_k(\frac{s}{r})}].$$

Since $\overline{T}_\epsilon^r \le T$ and using basic algebra, it is easy to check that there exists a constant $\xi \in [0, 1)$ such that

$$\mathbb{P}(A_r) \le 1 - (1 - \xi^r)^{rT} =: g(r). \tag{43}$$

We will now show that

$$\mathbb{P}(\cap_{r=1}^\infty \cup_{\tilde{r}=r}^\infty A_{\tilde{r}}) = 0. \tag{44}$$

In order to prove this, we need $\sum_{r=1}^\infty g(r) < \infty$. Noting that $\log(1 + x) = x + o(x)$ when $x$ is close to zero, and applying this fact to $\log((1 - \xi^r)^{rT})$, we obtain that $(1 - \xi^r)^{rT} = e^{-rT\xi^r + o(r\xi^r)}$ for large values of $r$. Using Taylor expansions and the fact that $\sum_{r=1}^\infty r\xi^r < \infty$, it follows that,

$$\begin{aligned} \sum_{r=1}^\infty g(r) &= \sum_{r=1}^\infty (1 - (1 - \xi^r)^{rT}) \\ &= \sum_{r=1}^\infty 1 - e^{-rT\xi^r + o(r\xi^r)} < \infty. \end{aligned}$$

Hence, we can apply Borel-Cantelli's lemma [19] and we obtain Equation (44).

We will now show that at time $T$ the weak fluid limit is almost surely equal to zero. From Equation (44) we obtain that, almost surely, event $A_r^c$ occurs when $r$ is large enough. Hence, only users in their best possible state are served in the interval $[0, r\overline{T}_\epsilon^r]$ when $r$ is large enough, i.e., $T_{k,n}^{BR,r}(t) = 0$, for $n \ne N_k$ and $t \le r\overline{T}_\epsilon^r$, for $r$ large enough, a.s.. Hence, for almost all $\omega$ we have that in the weak fluid limit presentation of Lemma IV.1 it holds that $\sum_{k=1}^K \tau_{k,N_k}^{BR}(t) = t$ and $\tau_{k,n}^{BR}(t) = 0$ for all $n \ne N_k$, $t \le \liminf_{r \to \infty} \overline{T}_\epsilon^r$. Therefore, for $t \le \liminf_{r \to \infty} \overline{T}_\epsilon^r$, we have that any weak fluid limit $y^{BR}(\cdot)$ satisfies

$$\begin{aligned} \sum_{k=1}^K \frac{y_k^{BR}(t)}{\mu_{k,N_k}} &= \sum_{k=1}^K \frac{x_k(0)}{\mu_{k,N_k}} + \sum_{k=1}^K \frac{\lambda_k}{\mu_{k,N_k}} t - \sum_{k=1}^K \tau_{k,N_k}^{BR}(t) \\ &= \sum_{k=1}^K \frac{x_k(0)}{\mu_{k,N_k}} - \left(1 - \sum_{k=1}^K \frac{\lambda_k}{\mu_{k,N_k}}\right) t. \end{aligned} \tag{45}$$

Let $T_\epsilon < \infty$ denote the moment that $\sum_{k=1}^K \frac{y_k^{BR}(t)}{\mu_{k,N_k}} = \epsilon$. For a given sample path $\omega$, let $r_k$ be the subsequence corresponding to $\liminf_{r \to \infty} \overline{T}_\epsilon^r$. By Lemma IV.1 we know that there exists a subsequence $r_{k_l}$ of $r_k$ such that $|\sum_k \frac{y^{BR}(\overline{T}_\epsilon^{r_{k_l}})}{\mu_{k,N_k}} - \sum_k \frac{Y^{BR,r_{k_l}}(\overline{T}_\epsilon^{r_{k_l}})}{\mu_{k,N_k}}| \leq \epsilon'$, for $\epsilon' > 0$ and $l$ large enough. Hence, if $\overline{T}_\epsilon^{r_{k_l}} \leq T_\epsilon$, we have that $T_\epsilon - \overline{T}_\epsilon^{r_{k_l}} \leq \frac{\epsilon'}{1-\sum \lambda_k/\mu_{k,N_k}}$, i.e., $\overline{T}_\epsilon^{r_{k_l}} \geq T_\epsilon - \frac{\epsilon'}{1-\sum \lambda_k/\mu_{k,N_k}}$. Since $\lim_{\epsilon' \downarrow 0} \liminf_{r \to \infty} \overline{T}_\epsilon^r \geq T_\epsilon$, the description (45) holds for all $t \leq T_\epsilon$. Now letting $\epsilon$ go to zero, then $T_\epsilon \to T$, hence Equation (45) holds for any $t \leq T$, and in particular $|y^f(T)| = 0$.

So we conclude that for almost all sample paths, any weak fluid limit converges to zero at time $T$, in particular the sequence corresponding to the liminf and the limsup. Hence, one can conclude that $\lim_{r \to \infty} Y_k^{f,r}(T) = y_k^f(T) = 0$, $k = 1, \ldots, K$, almost surely. Since $Y_k^{f,r}(T)$ converges almost surely (and therefore, also in probability) to $0$, we can use the same argument as in Theorem V.1, in order to conclude that the system will be stable under policy $f$, which finishes the proof of Theorem V.2.

Below we state the rate stability result.

**Lemma VIII.3.** *Assume $X(t)$ is rate stable, i.e., $\lim_{t \to \infty} X_i(t)/t = 0$ for all $i$. Then $\sum_{k=1}^{K} \frac{\lambda_k}{\mu_k} \leq 1$.*

**Proof:** We first derive rate conservation equations for the (per class) number of users. Using the martingale decomposition of the Markov chain, we obtain that $X_i(t)/t = x_i(0)/t + \sum_{s=1}^{t} \delta_i(X(s))/t + M_i(t)/t$, where $M_i$ is a square martingale such that $E[M_i(t)^2] \leq Ct$, with $C > 0$ a constant. Hence, $\frac{M(t)}{t} \to 0$, almost surely. So, if $\frac{X_i(t)}{t} \to 0$ then

$$\lim_{t \to \infty} \frac{1}{t} \sum_{s=1}^{t} \delta_i(X(s)) = 0. \tag{46}$$

Hence the time average of the Cesaro mean of the service $s_i(\cdot)$ dedicated to class $i$ is equal to $\lambda_i$. Given that we consider a system with capacity $1$, $\sum_{i=1}^{K} \frac{s_i(t)}{\mu_{i,N_i}} \leq 1, \forall t$, which, combined with the previous equation, gives that $\sum_{i=1}^{K} \frac{\lambda_i}{\mu_{i,N_i}} \leq 1$. $\qquad \square$